\documentclass{mem}
\usepackage{natbib}\usepackage{txfonts}\usepackage{balance}
\usepackage{graphicx}
\usepackage[a4paper,breaklinks,dvipdfm]{hyperref}
\idline{75}{1}
\begin{document}
\def\teff{$T\rm_{eff }$}
\def\kms{$\mathrm {km s}^{-1}$}
\def\msol{$M_{\odot}$}

\title{
Massive Be and Oe stars at low metallicity and long gamma ray bursts
}


\author{
C.\ Martayan\inst{1,2} 
\and J.\ Zorec\inst{3} 
\and D.\ Baade\inst{4} 
\and Y.\ Fr\'emat\inst{5} 
\and S.\ Ekstr\"om\inst{6}
\and J.\ Fabregat\inst{7}
          }

  \offprints{C. Martayan}
 
\institute{
European Organization for Astronomical Research in the Southern 
Hemisphere, Alonso de Cordova 3107, Vitacura, Santiago de Chile, Chile
\and
GEPI, Observatoire de Paris, CNRS, Universit\'e Paris Diderot, 5 place 
Jules Janssen, 92195 Meudon Cedex, France
\and
Institut d'Astrophysique de Paris, UMR7095, CNRS, Universit\'e Marie \& Pierre Curie, 
98bis Boulevard Arago 75014 Paris, France
\and
European Organisation for Astronomical Research in the Southern 
Hemisphere, Karl-Schwarzschild-Str.\ 2, 85748 Garching b.\ M\"unchen,
Germany
\and
Royal Observatory of Belgium, 3 avenue circulaire, 1180 Brussels, Belgium 
\and
Geneva Observatory, University of Geneva, Maillettes 51, 1290 Sauverny, Switzerland
\and
Observatorio Astron\'omico de Valencia, edifici 
Instituts d'investigaci\'o, 
Poligon la Coma, 46980 Paterna Valencia, Spain
\email{cmartaya@eso.org}
}

\authorrunning{Martayan et al.}

\titlerunning{Massive SMC Be/Oe stars and LGRBs}

\abstract{
According to recent theoretical studies, the progenitors of Long Gamma Ray Bursts should 
be very fast rotating stars, massive enough but not so for collapsing into a black hole.
In addition, recent observations seem to show that stars of about 20 solar masses could be 
at the origin of LGRBs. At low metallicity B-type stars rotate faster than at higher metallicity. 
We found with the ESO-WFI an occurrence of Be/Oe stars, that are quasi critical rotators, 3 to 5 
times larger in the SMC than in the Galaxy.   
According to our results, and using observational clues on the SMC WR stars, as well as the 
theoretical predictions of the characteristics must have the LGRB progenitors, 
we have identified the low metallicity massive Be/Oe stars as potential LGRB progenitors. 
To support this identification, the expected rates and the numbers of LGRB were then calculated 
and compared to the observed ones: 3 to 6 LGRBs were found in the local universe in 11 years 
while 8 were actually observed.
}
\maketitle{}

\section{Introduction}

The first section of this document deals with the Be phenomenon, their rates, their rotational velocities
in low metallicity environment.
The second section deals with the long gamma ray burst or type 2 bursts
and their links with the massive Be and Oe stars at low metallicity, 
according to the last theoretical developments
and observational facts.

\section{Be stars in low metallicity environment}
\label{section1}
The Be-phenomenon concerns the main sequence OBA-type stars, which spectrum has displayed at least
once in its life emission lines mainly in the hydrogen.
These lines come from a rotationally supported circumstellar disk 
formed by matter ejections of the central star.
Some information regarding the Be phenomenon can be found in \citet{porter03}.
Nowadays, we know that Be stars are rotating very fast, very close to
their critical rotational velocity but it is not clear whether an additional mechanism is needed
for ejecting the matter.
It was also reported that the Be-phenomenon seems to depend on the metallicity too \citep{maeder99}
and evolutionary status \citep[e.g.,][]{fabregat00,martayan07}.
A possibility for investigating these effects is the study of open clusters.
In the Milky Way a reference study is the one by \citet{mcs05}.

\subsection{Ratios of Be stars at low metallicity}
\label{Berates}
\begin{figure}[h!]
\includegraphics[angle=0, width=6.5cm]{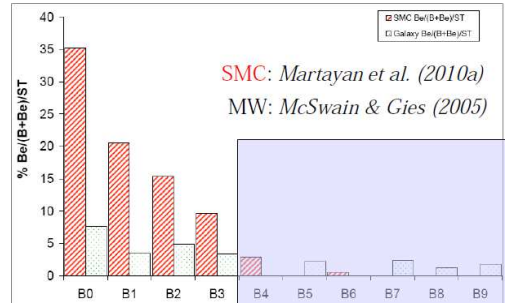}
\caption{\footnotesize
Ratios of Be stars to all B-type stars per spectral type categories.
The SMC sample is only complete till B3 spectral type.
The SMC ratios are compared to those from \citet{mcs05} in the Milky Way.
}
\label{fig1}
\end{figure}

\citet{maeder99} and \citet{wis07} found that the number of Be stars by open clusters
seems to increase with the decrease of the environment metallicity. However, the number of stars
and open clusters used is relatively small and it was necessary to improve the statistics 
by increasing the number of clusters, to quantify and compare the trends/ratios
between the Small Magellanic Cloud (a low metallicity galaxy) and the Milky Way.
\citet{martayan10a} did a spectroscopic survey of the SMC with the aim to find the emission-line 
stars and the Oe/Be/Ae stars. The ESO WFI \citep{baade99} in its slitless mode was 
used and about 3 million spectra were obtained.
The spectra of 4300 stars in SMC open clusters were extracted, analyzed, 
and the emission line stars found (for more details see \citet{martayan08, martayan10a}).
Then by cross-correlation with the OGLE catalogues \citep{udalski08} the stars were classified in spectral type.
With the freedom degrees constrained (age, metallicity, etc) the rates of Be stars to B type 
stars per spectral-type category were computed in the SMC and in the Galaxy with data of \citet{mcs05}.
The comparison of the Be to B stars per spectral type category rates is shown in Fig.\ \ref{fig1}.

The SMC sample is complete till spectral type B3. For spectral type ranging from B0 to B3 the 
rates in Fig.\ \ref{fig1} indicate that there are 3 to 5 times more Be stars in low metallicity environment 
than in our Galaxy.

\subsection{rotational velocities at low metallicity}

As shown in Sect.\ \ref{Berates}, the occurrence of the Be phenomenon is larger in lower metallicity environment.
It is explained by the fact that at lower metallicity O, B, A, and Be stars rotate faster than their counterparts
in higher metallicity environment \citep{keller04, martayan06, martayan07, hunter08}.
It is due to their lower mass-loss at low metallicity \citep{bouret03,vink07} that implies
a lower angular momentum loss \citep{maeder01}.

\begin{figure}[h!]
\includegraphics[angle=0, width=6.5cm]{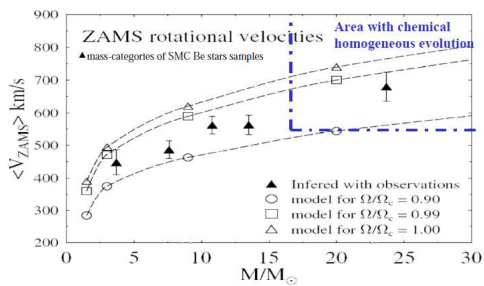}
\caption{\footnotesize
ZAMS rotational velocities of SMC Be and Oe stars compared to theoretical curves 
from the models of \citet{ekstrom08}.
In addition a box shows the location of stars that according to the theory \citep{brott11}
can follow the quasi chemically homogeneous evolution.
}
\label{fig2}
\end{figure}

With FLAMES observations \citet{martayan07} determined the rotational velocities of SMC Be, B stars.
They also determined their ZAMS rotational velocities taking into account the fast rotation effects 
such as the gravitational darkening \citep{fremat05, zorec05}.
The ZAMS rotational velocities of SMC Be stars are compared with theoretical models from \citet{ekstrom08}
in Fig.\ \ref{fig2}.
Their models seem to be able to reproduce properly the observations.

One can also note that SMC Be stars at their birth rotate very fast, with ZAMS rotational velocities exceeding
550 \kms for the most massive of them. However, for such very fast rotators, the stellar evolution could be modified
and instead of following a classical stellar evolution they could follow a quasi chemically homogeneous evolutionary 
as described in \citet{brott11}. In Fig.\ \ref{fig2}, a box with dashed lines shows the predicted area of stars
following this quasi chemically homogeneous evolution.

Therefore one can conclude that the massive Be and early Oe in low metallicity environment such as the SMC 
could follow the chemically homogeneous evolution. 
It is worth to notice that \citet{martins09} found SMC WR with properties that could be explained only
from the chemically homogeneous evolution.


\section{LGRBs and Oe/Be stars}
\label{section2}
This section deals with the type 2 burst also called long soft gamma ray burst (LGRB) 
because they are longer than 1-3s.
\citet{woosley93} proposed to explain the LGRB, the model of a massive fast rotating star collapsing
into a black-hole during SNIb,c explosion.
The energy for the explosion is provided by the potential
energy of the matter falling onto the collapsed stellar core. This remaining energy will not accrete 
onto the black hole that implies for the infalling matter to have enough angular momentum to remain in a
disk before accretion.

\subsection{Observational and theoretical facts}

\citet{iwamoto98,iwamoto00} support the idea that massive fast rotating stars are at the origin of the
LGRBs.
\citet{thone08} found that the LGRB \object{GRB060505} is hosted in a low-metallicity galaxy,
with a high star-formation rate. The LGRB came from a young environment (6
Myears) and from an object having about 32 \msol.
\citet{campana08} found that the LGRB060218 progenitor had an initial mass of 20 \msol.

Theoretically, the rotation seems to be a key point to understand the
appearance of GRBs \citep{woosley93,hirschi05,fryer07}. To keep a large amount of angular
momentum up to the last evolutionary phases before the collapse, GRBs progenitors should be massive
objects with low initial metallicities. According to \citet{yoon06} WR stars with 
metallicities $Z\!\lesssim0.002$ can be progenitors of GRBs.

Due to fast
rotation and the rotational mixing of chemical elements, massive stars can undergo
quasi-chemically homogeneous evolution to end up as helium WR stars satisfying the requirements for the
collapsar scenario \citep{yoon06}. \citet{yoon06} have calculated diagrams of LGRBs progenitors at different metallicities 
(including the SMC one) as a
function of their ZAMS rotational velocities and masses for magnetized massive stars following the
quasi-chemically homogeneous evolution. 

From their models, at low metallicity the WR phenomenon can appear in
stars having lower masses than those in the Milky Way.
We recall the observations by \citet{martins09}
of several SMC WR stars, whose evolutionary status and chemical properties can be understood if
they are fast rotators following this quasi-chemically homogeneous evolution.

According to the different criteria shown by the observations and the theory, the stars must be 
massive or of intermediate mass, fast rotator, formed in low metallicity environments.
Complying with these criteria and the diagrams of LGRB progenitors massive SMC Be and Oe stars
could become WR star and give a LGRB.

\subsection{Predicted rates of LGRBs from Oe/Be populations}
For testing the hypothesis of the SMC massive Be and Oe stars as LGRB progenitors, one can predict the 
rates of LGRBs based on that specific stellar population.
One can use the SMC as a test-bed galaxy, the number of OB stars is determined with the OGLE catalogues.
Then the number of the SMC Be/Oe stars is determined through the rates shown in Sect.\ \ref{Berates}, 
for more details, see also \citet{martayan10b}.

A base rate of LGRB/year is then obtained. However the gamma rays are strongly collimated.
One has to take into account the beaming angle to compute the probability to see a GRB.

The table\ \ref{LGRBrates} gives the LGRB base rates and the LGRB rates depending on the beaming angle considered
and for different mass category of stars (from B2e to O8e). According to the mass calibration of \citet{huang06}
the stars of spectral type B0 and higher (masses above 20 \msol) should be considered here.

\begin{table*}[h!]
\caption{Predicted LGRB rates}
\label{LGRBrates}
\begin{center}
\begin{tabular}{cllll}
\hline
\\
Mass category & LGRB base rate & Proba 5deg & Proba 10deg & Proba 15deg  \\
\hline
\\
B2e to O8e &  3.9-4.8 $\times 10^{-4}$ & 2.5-3.1 $\times 10^{-7}$ & 1.0-1.2 $\times 10^{-7}$ & 2.3-2.8 $\times 10^{-6}$ \\
B1e to O8e &  2.6-3.0 $\times 10^{-4}$ & 1.7-1.9 $\times 10^{-7}$ & 6.7-7.7 $\times 10^{-7}$ & 1.5-1.7 $\times 10^{-6}$ \\
\\
\hline
\\
B0e to O8e &  1.7-1.8 $\times 10^{-4}$ & 1.1-1.2 $\times 10^{-7}$ & 4.4-4.6 $\times 10^{-7}$ & 0.98-1.0 $\times 10^{-6}$ \\
O9e to O8e &  5.3-6.2 $\times 10^{-5}$ & 3.4-4.0 $\times 10^{-8}$ & 1.4-1.6 $\times 10^{-7}$ & 3.1-3.6 $\times 10^{-7}$ \\
     O8e   &  2.4-2.8 $\times 10^{-5}$ & 1.6-1.8 $\times 10^{-8}$ & 6.2-7.2 $\times 10^{-8}$ & 1.4-1.6 $\times 10^{-7}$ \\
\\
\hline
\end{tabular}
\end{center}
\end{table*}

However, one can take into account the LGRB phenomenon beaming angle distribution \citep{watson06}
instead of discrete angle values.
In such case, one can get:
$N_{\rm LGRB}^{\rm pred}\!\sim\!(2-5)\times10^{-7}$ LGRBs/yr/galaxy that can be compared to
the observed rate of LGRB:
$N_{\rm LGRB}^{\rm obs}\!\sim\!(0.2-3)\times10^{-7}$ LGRBs/yr/galaxy \citep{pod04, zhang04}.

\subsection{Predicted number of LGRBs in the local universe}
Another possibility is to predict the number of LGRBs in the local universe, i. e, at redshift$\le$0.2
for which the galaxy catalogue of \citet{skr06} is complete.
Up to redshift 0.5 the number of Im is about 17\% \citep{rocca07}.

One can get the predicted number of LGRBs in the local universe during 11 years (between 1998 and 2008)
using the previous predicted rates of LGRBs.
The result is shown in table\ \ref{LGRBnumbers} for numbers by mass category and discrete values
of beaming angles.

\begin{table*}[h!]
\caption{Predicted LGRB numbers in 11 years in the local universe (redshift$\le$0.2). }
\label{LGRBnumbers}
\begin{center}
\begin{tabular}{cccc}
\hline
\\
Mass category  & Number for angle=5\degr & Number for angle=10\degr & Number for angle=15\degr \\
\hline
\\
B2e to O8e &  3-4  & 11-14  & 25-31   \\
B1e to O8e &  2-2  & 7-9    & 16-19   \\
\\
\hline
\\
B0e to O8e &  1-1  & 5-5    & 11-11   \\
O9e to O8e &  0-0  & 2-2    & 3-4     \\
     O8e   &  0-0  & 1-1    & 2-2     \\
\\
\hline
\end{tabular}
\end{center}
\end{table*}

Ditto, one can also use the beaming angle distribution, in such case the predicted numbers of LGRBs are:
$N_{\rm LGRB}^{\rm pred}\!\sim\!3-6$ LGRBs (for mass categories starting from B0e to O8e).
This number could be compared to the observed number of LGRBs in the local universe in 11 years with data of the 
the GRBox from the University of California at 
Berkeley\footnote{see http://lyra.berkeley.edu/grbox/grbox.php} 
for years between 1998 and 2008, there are 8 LGRBs.

Therefore there is a relatively good agreement between our predictions and the observed rates/numbers 
of LGRBs in the local universe. 

\subsection{Binaries}
\label{binar}
Malesani et al. (this volume) report that several LGRBs occasionally occurred in 
``high'' metallicity environment. If confirmed it indicates that the usual models
of LGRBs, which need low metallicity environment cannot handle them.
Therefore, the binary models of LGRB creation \citep{cantiello07} could be of interest too.
Considering too the binary system as potential progenitor of LGRB, one could
multiply the rates we estimated above by 1/0.7, which gives a number of 9 LGRBs in the local universe 
(to be compared with the 8 LGRBs actually observed).

\subsection{Other biases}
Other possible biases than the binaries could be taken into account such as the missed GRBs by the satellites, the obscured GRBs not 
detected as GRB, etc. On the other side, one could also take into account the effect of the very fast rotating 
non emission line star, the so called Bn stars on the statistics.
For more details, the reader could consult the paper by \citet{martayan10b}.

\begin{acknowledgements}
C. M. acknowledges the useful comments by R. Hirschi and P. Vreeswijk.
\end{acknowledgements}

\bibliographystyle{aa}

\end{document}